\documentclass[pra,epsfig,rotate,superscriptaddress,showpacs]{revtex4}
\usepackage{graphicx}
\usepackage{epstopdf}
\usepackage{amsfonts}
\usepackage{amssymb}
\usepackage{amsmath}
\usepackage{subfigure}
\usepackage{prettyref}
\usepackage{float}
\usepackage{color,CJK}
\usepackage{amsmath}
\usepackage{amssymb}
\usepackage{esint}

\usepackage{braket}
\def\be{\begin{equation}}
\def\ee{\end{equation}}
\def\ba{\begin{eqnarray}}
\def\ea{\end{eqnarray}}

\begin{document}
%
%\begin{CJK}{UTF8}{gbsn}
\title
{The Non-Hermitian quantum mechanics and its canonical
structure}

\author{Qi Zhang}
\affiliation{College of Science, Zhejiang University of Technology, Hangzhou 310023, China}

\date{\today}
\begin{abstract}
The non-Hermitian Schr\"odinger equation is re-expressed generally in the form of Hamilton's canonical equation without any approximation. Its quantization called non-Hermitian quantum field theory is discussed. By virtue of the canonical equation, the theory of non-Hermitian quantum mechanics is totally reformulated, including the probability amplitudes of states, the expectations of operators, as well as the expressions of interaction terms. The conventional difficulties in non-Hermitian quantum mechanics are totally overcome by the reformulation.  Specifically, the imaginary parts the non-Hermitian eigenenergy and adiabatic geometric phase are actually unphysical, although they are mathematically perfect.
\end{abstract}
\pacs{03.65.-w,03.65.Vf}
%03.65.Vf: Phases: geometric; dynamic or topological
%Quantum mechanics, 03.65.-w
%45.20.Jj: Lagrangian and Hamiltonian mechanics

\maketitle
%\end{CJK}
\section{Introduction}

The microscopic world is naturally described by the quantum mechanics (QM) where the Hamiltonians and observables are all Hermitian operators. Alternatively, researchers study the non-Hermitian QM theoretically to explore the $\mathcal{PT}$-symmetric mechanics~\cite{Bender}. The non-Hermitian mechanics arises in special representation of bosonic Bogoliubov quasiparticle system as well as in the open quantum systems~\cite{ZhangNJP,njp}.
There have been tremendous interests in non-Hermitian systems both theoretically ~\cite{Bender,Bender2,Bender3,Mostafazadeh,Berry,Longhi,West,Mostafazadeh2,Bender4,Bender5,Wang,chen,Fu,chen1,chen2,Rego} and experimentally~\cite{e1,e2,e3,e4,e5,e6,e7,e8,e9,e10,e11,e12,e13,J1,J3,J4,J5}.

Just as the difficulty of negative probability (energy) of relativistic quantum mechanics before the establishment of quantum field theory, fundamental difficulties naturally arise about the non-Hermitian QM: (i) since the right eigenvectors and the corresponding left ones of a non-Hermitian Hamiltonian do not coincide, it constitutes an open question which state the system is actually on when it is said to be on an eigenstate; (ii) will the complex eigenenergies really result in the non-conservation of probability amplitude? (iii) the complex Berry phase arises as a consequence of the biorthogonal eigenvectors~\cite{ZhangPRA,GarrisonPLA,GePRA}, but other researchers take the argument that the geometric phase must be real as an axiom~\cite{SamuelPRL,Gong}.

In this paper, we generalize the canonical formulation of the non-Hermitian Schr\"odinger equation that had been explored for a specific case~\cite{Rego}. Based on the canonical structure, namely the non-Hermitian field theory, it is found that most of the formulae in conventional QM cannot apply to the non-Hermitian QM. The reformulation of non-Hermitian QM overcomes the fundamental difficulties in non-Hermitian QM: (i) unlike QM, a non-Hermitian state is no longer a probability wave and the modified expression for the probability amplitude displays the conservation of the occupation number even when the eigenenergies are complex; (ii) the imaginary part of an eigenenergy and that of a geometric phase,  although mathematically perfect, is physically meaningless; (iii) the expectation value of an operator in non-Hermitian QM assumes the very different expression from the conventional QM; (iv) an intrinsic difficulty about probability occupation exists in non-Hermitian QM, which vanishes after the second-quantization.

\section{The canonical framework of non-Hermitian QM}

Consider the dynamics of a non-Hermitian system, which is given by the Schr\"odinger equation
\begin{equation} \label{Schro}
\text{i}\hbar\frac{\partial}{\partial t}|\psi\rangle=h|\psi\rangle\,.
\end{equation}
which resembles the conventional QM but with $h$ a non-Hermitian operator. As $h$ is not Hermitian, the norm $\braket{\psi(t)|\psi(t)}$ is not conserved during the dynamical evolution.
When a non-Hermitian Hamiltonian $h$ is diagonalizable, it has two sets of eigenvectors
$|a_j\rangle$ and $|b_j\rangle$ satisfying~\cite{F3},
\begin{equation} \label{right-left}
h|a_j\rangle=E_j|a_j\rangle,  \quad
\langle b_j|h=\langle b_j|E_j.
\end{equation}
They are  biorthonormal, $\langle b_i|a_j\rangle=\delta_{ij}$, and complete,
\begin{equation}  \label{right-left2}
\sum_j |a_j\rangle\langle b_j|=1\,.
\end{equation}
Usually $|a_j\rangle$ and $|b_j\rangle$ are called respectively the right eigenvectors and left eigenvectors. As $h^\dag\neq h$, it is obvious that $|a_j\rangle\neq|b_j\rangle$.

Two fundamental questions arise naturally for the non-Hermitian QM: (i) as $\braket{\psi(t)|\psi(t)}$ is not conserved, is it still the right expression for the overall probability? (ii) when the non-Hermitian system is on an eigenstate of $h$ with eigenvalue $E_j$, which state between $|a_j\rangle$ and $\langle b_j|$ is the system actually on? To answer the questions, we will try to reformulate the non-Hermitian QM by the canonical theory. Next the general form of canonical structure for the non-Hermitian QM is explored, which, for a very specific form of Hamiltonian, had been considered in~\cite{Rego}.

\subsection{The non-Hermitian system of discrete levels}

For clarity we first consider the countable-level system where the wavefunction has $n$ components and thus $h$ a $n\times n$ non-Hermitian matrix. The wave function $|\psi\rangle$ in Eq.~(\ref{Schro}) can then be expressed as $|\psi\rangle=(\psi_1,\psi_2,\ldots,\psi_n)^T$. The state of a system should be represented by the canonical variables, so $q_j=\text{i}\hbar\psi_j$ can be taken naturally as canonical variables.
Different from the Hermitian case, it is evident that, as $h\neq h^\dag$, $(\text i\hbar\psi_j,\psi_j^*)$ cannot constitute a pair of canonical variables~\cite{Rego}. Therefore, we first introduce $n$ variables denoted $\bar{\phi}_1,\bar{\phi}_2,\ldots,\bar{\phi}_n$, with $q_k=\text{i}\hbar\psi_k, p_k=\bar{\phi}_k$ form a pair of canonical variables, and then try to find their concrete expressions.

To let the Schr\"odinger equation (\ref{Schro}) be equivalent with the Hamilton's canonical equation,
\begin{equation} \label{FH}
\frac{d(\text{i}\hbar\psi_k)}{dt}=\frac{\partial\mathcal{H}}{\partial\bar{\phi}_k},  \quad
\frac{d(\bar{\phi}_k)}{dt}=-\frac{\partial\mathcal{H}}{\partial(\text{i}\hbar\psi_k)},
\end{equation}
the field Hamiltonian $\mathcal{H}$ for the canonical equation assumes,
\begin{equation} \label{3}
\mathcal{H}(\bar{\phi}_1,\ldots,\bar{\phi}_n,\text{i}\hbar\psi_1,\ldots,\text{i}\hbar\psi_n)=\langle\bar{\phi}|h|\psi\rangle.
\end{equation}
According to the expression (\ref{3}), the first half of the Hamilton's canonical equations for  $q_k=\text{i}\hbar\psi_k$ are satisfied apparently. Then finding $\bar{\phi}_j$s constitutes the main task. It is worth noting that the arbitrarily defined canonical momenta $p_k=\bar{\phi}_k$ will extend the dimension of the phase space from $2n$ to $4n$, violating the principle of canonical dynamics.
To find the canonical momenta satisfying the canonical equation,
we take advantage of the biorthonormal eigenstates of $h$ as given by Eq.~(\ref{right-left}).
We can first expand the classical field $|\psi\rangle=(\psi_1,\psi_2,\ldots,\psi_n)^T$ in terms of the right vectors,
\begin{equation} \label{DC}
|\psi\rangle=\sum_{j=1}^n c_j|a_j\rangle.
\end{equation}
With the expansion coefficients $c_j$s thus derived and the left eigenvectors $\langle b_j|$, another classical field $\langle\bar{\phi}|$ is constituted as,
\begin{equation} \label{MOMENTA}
\langle\bar{\phi}|=\sum_{j=1}^n\bar{c}_j\langle b_j| ,
\end{equation}
where $\bar{c}_j$ is set to be numerically equal to,
\begin{equation} \label{relationI}
\bar{c}_j=\frac{|C_j|^2}{c_j},
\end{equation}
with $|C_j|^2\equiv C_j^*C_j$ for $j=1,2,\ldots,n$ are assumed to be $n$ constants. One can check readily that the components of $\langle\bar{\phi}|=(\bar{\phi}_1,\bar{\phi}_2,\ldots,\bar{\phi}_n)$ defined in (\ref{MOMENTA}) satisfy the other half of the canonical equation shown in (\ref{FH}). For a given Hamiltonian $h$, $|a_j\rangle$ and $|b_j\rangle$ are known, the canonical momenta defined by (\ref{DC}) and (\ref{MOMENTA}) will not result in any redundant variables, in accord with the principle of canonical dynamics.

To clarify the canonical equation, one may carry out a time-independent canonical transformation
\begin{equation} \nonumber
(\bar{\phi}_1,\ldots,\bar{\phi}_n,\text{i}\hbar\psi_1,\ldots,\text{i}\hbar\psi_n)\rightarrow(\bar{c}_1,\ldots,\bar{c}_n,\text{i}\hbar c_1,\ldots,\text{i}\hbar c_n),
\end{equation}
with the Hamiltonian $\mathcal{H}$ numerically unchanged. The canonical equation now reads,
\begin{equation} \label{CAN}
\frac{d(\text{i}\hbar c_j)}{dt}=\frac{\partial \mathcal{H}}{\partial \bar{c}_j},\quad\frac{d \bar{c}_j}{dt}=-\frac{\partial \mathcal{H}}{\partial(\text{i}\hbar c_j)},
\end{equation}
with $\mathcal{H}$ expressed in terms of $c_j$ and $\bar{c}_j$ as
\begin{equation} \label{14}
\mathcal{H}=\sum_{j=1}^n E_j\bar{c}_jc_j.
\end{equation}
According to the canonical equation Eq.~(\ref{CAN}) and the relation given by (\ref{relationI}), the transformed canonical variables $c_j$ and $\bar{c}_j$ evolve as
\begin{equation} \label{EV1}
c_j(t)=D_je^{-\frac{\text{i}}{\hbar}E_jt},\quad \bar{c}_j(t)=\bar{D}_je^{\frac{\text{i}}{\hbar}E_jt},
\end{equation}
where $\bar{D}_jD_j=|C_j|^2$ are constant. According to Eq.~(\ref{EV1}), the (complex) field energy $\mathcal{H}$ as shown in (\ref{3}) and (\ref{14}) is conserved even though $E_j$s may be complex. In addition, the relation (\ref{relationI}) holds constantly during the canonical dyanmics, demonstrating the self-consistency of the framework.

The Lagrangian of the non-Hermitian field can then be expressed as,
\begin{equation}
\mathcal{L}={\text i}\hbar\sum_{j=1}^n\bar{\phi}_j\dot{\psi}_j-\langle\bar{\phi}|h|\psi\rangle,
\end{equation}
where the Lagrangian equation
\begin{equation}
\frac{d}{dt}\frac{\partial \mathcal{L}}{\partial \dot{\psi}_j}-\frac{\partial \mathcal{L}}{\partial \psi_j}=0
\end{equation}
is found to be equivalent to the canonical equation (\ref{FH}).

\subsection{The non-Hermitian quantum system of continuous levels}

From a discrete field, it is possible to go to a continuous field when the number of degrees of freedom becomes noncountably infinite. For clarity, we adopt the coordinate representation for discussion by denoting $|\mathbf{x}\rangle$ the coordinate eigenvector with the eigenvalue $\mathbf{x}$, i.e., $\hat{x}|\mathbf{x}\rangle=\mathbf{x}|\mathbf{x}\rangle$ with $\hat{x}$ being the coordinate operator.
As in the discrete case, the eigenfunction assumes the biorthonormal form as shown in Eq.~(\ref{right-left}),
but with the $j$th right eigenvector $a_j(\mathbf{x})=\langle\mathbf{x}|a_j\rangle$ and left one $b_j(\mathbf{x})=\langle\mathbf{x}|b_j\rangle$ being continuous functions of coordinate $\mathbf{x}$. The biorthonormal condition and complete condition now read,
\begin{equation} \label{clr2}
\int b_j(\mathbf{x})^*a_k(\mathbf{x}) d\mathbf{x}=\delta_{jk}, \quad \sum_j a_j(\mathbf{x})b_j(\mathbf{x'})^* =\delta(\mathbf{x}-\mathbf{x}').
\end{equation}
In terms of the wave function $\psi(\mathbf{x})=\langle\mathbf{x}|\psi\rangle$, the Schr\"odinger equation reads,
\begin{equation} \label{NQM}
\text{i}\hbar\frac{\partial}{\partial t}\psi(\mathbf{x})=\hat{h}\psi(\mathbf{x}),
\end{equation}
with $\hat{h}$ being the Hamiltonian operator in coordinate representation. The relation between $\hat{h}$ and the matrix element of $h$ is $\langle \mathbf{x}'|h|\mathbf{x}\rangle=\hat{h}(\hat{x},-\text i\hbar\partial/\partial \mathbf{x})\delta(\mathbf{x}-\mathbf{x}')$.
The canonical variables can be found in the same way as in the countable-level case: (i) expanding $\psi(\mathbf{x})$ by
\begin{equation} \label{6}
\psi(\mathbf{x})=\sum_j c_j a_j(\mathbf{x});
\end{equation}
(ii) with $c_j$s and left eigenvectors constructing the canonical conjugate field
\begin{equation}  \label{7}
\bar{\phi}(\mathbf{x})=\langle\bar{\phi}|\mathbf{x}\rangle=\sum_j \bar{c}_j b_j^*(\mathbf{x}),
\end{equation}
with     \begin{equation}      \label{8}
\bar{c}_jc_j=|C_j|^2.  \end{equation}
As can be proven, the fields $\bar{\phi}(\mathbf{x})$ and $\psi(\mathbf{x})$ satisfy the canonical equation,
\begin{equation} \label{cFH}
\frac{d(\text{i}\hbar\psi_k)}{dt}=\frac{\partial\mathbb{H}}{\partial\bar{\phi}_k}, \quad
\frac{d(\bar{\phi}_k)}{dt}=-\frac{\partial\mathbb{H}}{\partial(\text{i}\hbar\psi_k)},
\end{equation}
with Hamiltonian density $\mathbb{H}(\mathbf{x})$ given by,
\begin{equation} \label{Hdensity}
\mathbb{H}=\bar{\phi}(\mathbf{x})\hat{h}\psi(\mathbf{x}).
\end{equation}
The Lagrangian density of the field then reads,
\begin{equation}
\mathbb{L}(\mathbf{x})={\text i}\hbar\bar{\phi}(\mathbf{x})\dot{\psi}(\mathbf{x})-\bar{\phi}(\mathbf{x})\hat{h}\psi(\mathbf{x}).
\end{equation}
The equation of motion satisfies the Euler-Lagrange equation for continuous field,
\begin{equation}
\frac{\partial\mathbb{L}}{\partial\psi}
-\nabla\cdot\frac{\partial\mathbb{L}}{\partial(\nabla\psi)}
-\frac{\partial}{\partial t}\left(\frac{\partial\mathbb{L}}{\partial\dot{\psi}}\right)=0.
\end{equation}

\section{Re-establishment of fundamental equations in non-Hermitian QM}

\subsection{The probability wave and probability conservation}

Now we discuss the significance of the canonical equation in reformulating the non-Hermitian QM and resolving the fundamental difficulties in it. According to the canonical equation of classical non-Hermitian field as shown in Eqs.~(\ref{FH}) and (\ref{cFH}), the overall field Hamiltonian $\mathcal{H}=\int \mathbb{H} d\mathbf{x}$ is conserved during temporal evolution, amounting to that, according to Eq.~(\ref{14}), $\bar{c}_jc_j$ is the occupation number of $|\psi\rangle$ on the $j$th energy eigenstate. As such, the overall probability for a right state $|\psi\rangle$ is {\it not} $\langle\psi|\psi\rangle$ but $\langle\bar{\phi}|\psi\rangle$, as shown clearly by Eqs.~(\ref{right-left}), (\ref{right-left2}), (\ref{DC}) and (\ref{MOMENTA}). The normalization condition for the non-Hermitian QM is then,
\begin{equation} \label{normalizationCC}
\langle\bar{\phi}|\psi\rangle=1.
\end{equation}
It is noteworthy that this modified normalization condition had been naturally applied in Bosonic Bogoliubov quasi-particle systems~\cite{ZhangNJP,ZhangNiu} for the neatness of formulae but without a systematic proof.
When the eigenenergies $E_j$s are complex for the non-Hermitian Hamiltonian, such as in the $\mathcal{PT}$-broken region of the $\mathcal{PT}$-symmetric mechanics~\cite{Bender}, it is usually considered that the wave function will enhance or decay exponentially with time, resulting in the nonconservation of the occupation number.
However, with the biorthogonal completeness condition (\ref{right-left2}) and the normalization (\ref{normalizationCC}) the correct physical occupation number is,
\begin{equation} \label{PROBA}
\sum_j \langle\bar{\phi}|a_j\rangle\langle b_j|\psi\rangle=\sum_j\bar{c}_jc_j=1,
\end{equation}
meaning the overall probability is conserved. The decaying of right/left eigenstate when the eigenenergies become complex is mathematical rather than physical. According to Eq.~(\ref{EV1}), the occupation on an energy eigenstate is also conserved.

Employing the completeness condition of coordinate eigenvectors $\int|\mathbf{x}\rangle\langle\mathbf{x}|d\mathbf{x}=1$, we get,
\begin{equation}
\int\langle\bar{\phi}|\mathbf{x}\rangle\langle\mathbf{x}|\psi\rangle d\mathbf{x}=\int\bar{\phi}(\mathbf{x})\psi(\mathbf{x})d\mathbf{x}=1,
\end{equation}
which means that $\bar{\phi}(\mathbf{x})\psi(\mathbf{x})$ is the real probability of finding a non-Hermitian particle of state $\psi(\mathbf{x})$ at $\mathbf{x}$. It is noteworthy that the probability $\bar{\phi}(\mathbf{x})\psi(\mathbf{x})$ may be negative for some $|a_j\rangle$ and $|b_j\rangle$, which can be solved, as will be seen, by the second-quantization.

\subsection{The expectations of operators}

With the probability expressed as $\bar{\phi}(\mathbf{x})\psi(\mathbf{x})$, the expectation value of coordinate for the right state $|\psi\rangle$ is derived,
\begin{eqnarray} \nonumber
\langle\bar{\phi}|\hat{x}|\psi\rangle&=&\int\int\langle\bar{\phi}|\mathbf{x}\rangle\langle\mathbf{x}|\hat{x}|\mathbf{x}'\rangle\langle\mathbf{x}'|\psi\rangle d\mathbf{x}d\mathbf{x}'\\
&=&\int \mathbf{x}\bar{\phi}(\mathbf{x})\psi(\mathbf{x})d\mathbf{x}.
\end{eqnarray}
In the very same manner, the occupation in any representation, denoted $\mathcal{O}$, is $\bar{\phi}(\mathcal{O})\psi(\mathcal{O})$. The expectation of the operator $\hat{\mathcal{O}}$ at the right state $|\psi\rangle$ is,
\begin{equation} \label{OOO}
\langle\hat{\mathcal{O}}\rangle=\langle\bar{\phi}|\hat{\mathcal{O}}|\psi\rangle.
\end{equation}
Specifically, according to Eq.~(\ref{OOO}), the energy expectation of state $|\psi\rangle$ shall be $\langle\bar{\phi}|h|\psi\rangle=\mathcal{H}$, which is nothing but the overall Hamiltonian of non-Hermitian field.

According to the conventional nonrelativistic QM, to get the information of a wave function, we may detect a physical variable, say $\mathbf{x}$, repeatedly.  The expressions for the probability wave $\bar{\phi}(\mathbf{x})\psi(\mathbf{x})$ and the expectation of operator $\langle\hat{\mathcal{O}}\rangle$ amount to that the non-Hermitian system, at any time $t$, is located simultaneously at the right state $|\psi\rangle$ and the corresponding left one $\langle\bar{\phi}|$. In fact, the left state and the right one can be considered as the same state from the viewpoint of covariant and contravariant vectors~\cite{ZhangPRA}.

\subsection{The meaning of complex geometric phase}

In studying the Berry-like phase in non-Hermitian system, parameters $\mathbf{R}$ of the Hamiltonian $h(\mathbf{R})$ are allowed to change slowly. It had been proven from adiabaticity that the instantaneous $j$th eigenstate evolves as (with the initial state on the $j$th energy eigenstate)~\cite{ZhangPRA,GarrisonPLA,GePRA},
\begin{equation} \label{e26}
|\psi(\mathbf{R})\rangle=|a_j(\mathbf{R})\rangle e^{-{\text i}\frac{\int E_j(\mathbf{R}) dt}{\hbar}}e^{{\text i}\beta_j}\equiv c_j|a_j(\mathbf{R})\rangle \,,
\end{equation}
where $\beta_j$ is the gauge-dependent geometric phase and $-\int E_j(\mathbf{R}) dt/\hbar$ the dynamical phase. For a closed path $\beta_j$ becomes the gauge-independent Berry phase.
The Berry connection is found to be associated with both right eigenvector and the corresponding left one~\cite{ZhangPRA,GarrisonPLA,GePRA},
\begin{equation} \label{Berry-conn}
\mathbf{A}_j=\frac{\partial \beta_j}{\partial \mathbf{R}}=\text{i}\langle b_j(\mathbf{R})|\frac{\partial}{\partial \mathbf{R}}|a_j(\mathbf{R})\rangle.
\end{equation}
The Berry curvature in three dimensional parameter space therefore takes the following form~\cite{GarrisonPLA},
\begin{equation} \label{Gcur}
\mathbf{B}_j=\nabla\times\mathbf{A}_j=\text{i}\langle\nabla b_j|\times|\nabla a_j\rangle,
\end{equation}
As $|a_j\rangle\neq|b_j\rangle$, the Berry connection and Berry curvature are generally not real for non-Hermitian systems. Will this complex geometric phase result in the geometric decreasing or increasing of particles in non-Hermitian QM, as stated in Ref.~\cite{GarrisonPLA}? To answer this question, we employ the fact that the occupation on $j$th eigenstate is not $c_j^*c_j$ but $\bar{c}_jc_j$. Evidently we should also consider the ``left" Schr\"odinger equation, $-\text{i}\hbar\frac{\partial}{\partial t}\langle\bar{\phi}|=\langle\bar{\phi}|h$ and calculate the geometric phase for the left eigenstate, denoted $\bar{\beta}_j$,
\begin{equation}
|\phi(\mathbf{R})\rangle=|b_j(\mathbf{R})\rangle e^{-{\text i}\frac{\int E_j^*(\mathbf{R}) dt}{\hbar}}e^{{\text i}\bar{\beta}_j}\equiv \bar{c}_j^*|b_j(\mathbf{R})\rangle \,,
\end{equation}
where the left state dynamical phase $-\int E_j^*(\mathbf{R}) dt/\hbar$ results from the eigenfunction $h^\dag|b_j\rangle=E_j^*|b_j\rangle$. The Berry connection for the left eigenstate can be worked out directly,
\begin{equation} \label{Berry-conn2222}
\mathbf{\bar{A}}_j=\frac{\partial \bar{\beta}_j}{\partial \mathbf{R}}=\text{i}\langle a_j(\mathbf{R})|\frac{\partial}{\partial \mathbf{R}}|b_j(\mathbf{R})\rangle.
\end{equation}
By comparing Eqs.~(\ref{Berry-conn}) and (\ref{Berry-conn2222}), it is evident from $\langle b_i|a_j\rangle=\delta_{ij}$  that $\mathbf{\bar{A}}_j=\mathbf{A}_j^*$ and thus $\bar{\beta}_j=\beta_j^*$.
After the adiabatic evolution, the probability to find the $j$the energy eigenstate is,
\begin{equation} \label{BERRRRR}
\bar{c}_jc_j=\left[e^{-{\text i}\frac{\int E_j^*(\mathbf{R}) dt}{\hbar}}e^{{\text i}\bar{\beta}_j}\right]^*e^{-{\text i}\frac{\int E_j(\mathbf{R}) dt}{\hbar}}e^{{\text i}\beta_j}=1,
\end{equation}
meaning that the physical decaying or enhancing of the wave function is fictitious in the non-Hermitian QM. This fact shows that the imaginary part of the geometric phase is merely a mathematical result. However, the real part standing for the phase factor can be really detected by the inference experiment. It is worth noting that this real geometric phase is still different from that in the previous study~\cite{SamuelPRL,Gong}, where the expression for the Berry connection is only associate with the right eigenstate.

The derivation of geometric phase is equivalent to the Hannay's angle accompanying the adiabaticity of the classical adiabatic evolution regarding the Hamiltonian $\mathcal{H}(\mathbf{R})=\langle\bar{\phi}(\mathbf{R})|h(\mathbf{R})|\psi(\mathbf{R})\rangle$. The canonical equation as shown in Eq.~(\ref{FH}) gives an integrable system with the actions $I_k$s,
\begin{equation}
I_k=\frac{\text i\hbar}{2\pi}\oint \bar{\phi}_k d\psi_k
\end{equation}
being the adiabatic invariants. As can be proven, the calculation of Hannay's angle regarding this adiabatic evolution just gives rise to the Berry phase in the non-Hermitian QM.

\section{Quantization}

The arbitrary constants $|C_j|^2$ given by Eq.~(\ref{relationI}) display an intrinsic ambiguity of non-Hermitian QM, i.e., for a certain right state $|\psi\rangle$, the occupation on the $k$th eigenstate $\bar{c}_kc_k$ can be set arbitrarily. In other words, they depend on the gauge transformation $|a_j'\rangle=f|a_j\rangle, \; \langle b_{j}'|=\frac{1}{f}\langle b_j|$, with $f$ a complex number ($|f|\neq1$ and $f\in GL(1,\mathbb{C})$).
This gauge difficulty, together with the problem of non-positive definite probability, can be overcome by quantizing the canonical equation (\ref{FH}). Promoting the canonical variables to operators, from the principle of canonical quantization it is evident that,
\begin{eqnarray} \label{commutation} \nonumber
&[\hat{\bar{\phi}}_j,\hat{\psi}_k]_{\mp}=\delta_{jk},\; [\hat{\bar{c}}_j,\hat{c}_k]_{\mp}=\delta_{jk},\; [\hat{C}_j^\dag,\hat{C}_k]_{\mp}=\delta_{jk},\\
&[\hat{\psi}_j,\hat{\psi}_k]_{\mp}=[\hat{\bar{\phi}}_j,\hat{\bar{\phi}}_k]_{\mp}=0,
\end{eqnarray}
with $[\ldots]_{\mp}$ standing for the commutator and anticommutator, respectively, meaning that the field can be quantized into either bosons or fermions. The third (anti)commutator in (\ref{commutation}) stems from the relation $\bar{c}_kc_k=C_k^*C_k$ before the quantization.
To find the elementary excitations, the quantized Hamiltonian is written by employing the relations (\ref{right-left}), (\ref{right-left2}), (\ref{DC}), (\ref{MOMENTA}), (\ref{relationI}), (\ref{6}), (\ref{7}) and (\ref{8}),
\begin{equation} \label{SECONDH}
\hat{\mathcal{H}}=\sum_{j=1}^n E_j \hat{\bar{c}}_j \hat{c}_j=\sum_{j=1}^n E_j \hat{C}_j^\dag \hat{C}_j.
\end{equation}
With the (anti)commutation relation (\ref{commutation}), the complex $E_j$s, which are also the eigenenergies of the original non-Hermitian system with Hamiltonian $h$, are found to be the energies of elementary excitations of the non-Hermitian field.
The quantization for the continuous-level system is straightforward,
\begin{eqnarray} \label{commutation2} \nonumber
&[\hat{\bar{\phi}}(\mathbf{x}),\hat{\psi}(\mathbf{x}')]_{\mp}=\delta(\mathbf{x}-\mathbf{x}'),\; [\hat{\bar{c}}_j,\hat{c}_k]_{\mp}=\delta_{jk},\\
&[\hat{\psi}(\mathbf{x}),\hat{\psi}(\mathbf{x}')]=[\hat{\bar{\phi}}(\mathbf{x}),\hat{\bar{\phi}}(\mathbf{x}')]=0
\end{eqnarray}
In the second-quantization Hamiltonian (\ref{SECONDH}), $\hat{C}_j^\dag \hat{C}_j$ ($\hat{\bar{\phi}}(\mathbf{x})\hat{\psi}(\mathbf{x})$) becomes the particle number operators rather than the gauge-dependent arbitrary constant (non-positive definite probability). Thus the difficulty of negative or gauge-dependent occupation number in non-Hermitian QM disappears naturally after quantizing the non-Hermitian field.

\section{The interaction in non-Hermitian QM}

We eventually discuss the basic form of two-body interaction term in non-Hermitian QM. For concreteness, the Gross-Pitaevskii-like equation is taken as an example. According to occupation number $\bar{\phi}(\mathbf{x})\psi(\mathbf{x})$ of state $\psi(\mathbf{x})$ at $\mathbf{x}$, the Gross-Pitaevskii equation is recognized to be,
\begin{equation} \label{INTER_HH}
{\text i}\hbar\frac{\partial}{\partial t}\psi=\left[-\frac{\hbar^2}{2m}\frac{\partial^2}{\partial \mathbf{x}^2}+V(\mathbf{x})+c\bar{\phi}(\mathbf{x})\psi(\mathbf{x})\right]\psi,
\end{equation}
with $V(\mathbf{x})$ being a complex potential and $c$ proportional to the s-wave scattering length. The time-independent eigenfunction can then be written directly. The left wave function $\bar{\phi}(\mathbf{x})$ is determined by a self-consistent way: (i) for a given $\psi(\mathbf{x})$, the Hamiltonian operator $\hat{h}$ is totally determined by a left state initially taken as $\bar{\phi}(\mathbf{x})$; (ii) derive all the left and right eigenstates $a(\mathbf{x})$, $b(\mathbf{x})$ of $\hat{h}$; (iii) find the new left state $\bar{\phi}(\mathbf{x})$ according to Eqs.~(\ref{6}), (\ref{7}) and (\ref{8}) within a certain gauge $|C_j|^2$; (iv) the left state thus derived should coincide with that assumed initially. Numerically the left state corresponding to a given right state can be obtained iteratively.

The quantization of non-Hermitian Gross-Pitaevskii dynamics is straightforward. According to Eq.~(\ref{Hdensity}) and $\hat{h}$ shown in Eq.~(\ref{INTER_HH}), the field Hamiltonian reads,
\begin{equation}
\mathcal{H}=\int\left[\bar{\phi}(\mathbf{x})\hat{h}'\psi(\mathbf{x})+\frac{1}{2}\lambda \bar{\phi}(\mathbf{x})\bar{\phi}(\mathbf{x})\psi(\mathbf{x})\psi(\mathbf{x})\right]d\mathbf{x},
\end{equation}
where $\hat{h}'=-\frac{\hbar^2}{2m}\frac{\partial^2}{\partial \mathbf{x}^2}+V(\mathbf{x})$ is the Hamiltonian excluding the interaction term and $c=N\lambda$ (with $N$ being the total particle number, which, according to our result, is conserved even if the eigenenergies are complex). The second-quantized Hamiltonian is obtained directly by promoting the field functions to operators with the (anti)commutation relation given by Eq.~(\ref{commutation2}),
\begin{equation}
\hat{\mathcal{H}}=\int\left[\hat{\bar{\phi}}(\mathbf{x})\hat{h}'\hat{\psi}(\mathbf{x})+\frac{1}{2}\lambda \hat{\bar{\phi}}(\mathbf{x})\hat{\bar{\phi}}(\mathbf{x})\hat{\psi}(\mathbf{x})\hat{\psi}(\mathbf{x})\right]d\mathbf{x}.
\end{equation}
When encountering an action at a distance $U(\mathbf{x},\mathbf{x}')$ in the nonrelativistic theory, the second quantized Hamiltonian will be,
\begin{equation}
\hat{\mathcal{H}}=\int\hat{\bar{\phi}}(\mathbf{x})\hat{h}'\hat{\psi}(\mathbf{x})d\mathbf{x}+\int\int\frac{1}{2} \hat{\bar{\phi}}(\mathbf{x})\hat{\bar{\phi}}(\mathbf{x}')U(\mathbf{x},\mathbf{x}') \hat{\psi}(\mathbf{x}')\hat{\psi}(\mathbf{x})d\mathbf{x}d\mathbf{x}'.
\end{equation}
In some condensed matter systems, the excitations may be described by the interacting non-Hermitian Hamiltonian such as the excitations in Nonequilibrium Bose-Einstein condensate of exciton polaritons~\cite{WPRL2007}.

\section{Summary}

To summarize, we have established the canonical framework for the nonrelativistic non-Hermitian QM. The fundamental difficulties in non-Hermitian QM are found to be overcome by the canonical structure and its quantization. We hope the current work will stimulate future studies in non-Hermitian QM. Besides, the exploration of non-Hermitian quantum statistics or introducing the non-Hermiticity to relativistic field theory should also be of considerable interest.

%This work was supported by the The National Key R\&D Program of China (Grants No.~2017YFA0303302, No.~2018YFA0305602).
%Besides, the exploration of non-Hermitian quantum statistics or introducing the non-Hermiticity to relativistic field theory should also be of considerable interest.

%\section{acknowledgement}
%This work was supported by the The National Key Research and Development Program of China (Grants No.~2017YFA0303302, No.~2018YFA030562) and the National Natural Science Foundation of China (Grants No.~11334001 and No.~11429402).

\end{document}